\documentclass{PoS}

\title{Radiative contribution to the effective potential in a composite Higgs model}

\ShortTitle{Radiative contribution to the effective potential in a composite Higgs model}

\author{T. A. DeGrand,$^a$ M. Golterman,$^b$ W. I. Jay,$^a$ E. T. Neil,$^{ac}$
Y. Shamir$^d$ and \speaker{B.~Svetitsky}$^d$\\
\llap{$^a$}Department of Physics, University of Colorado, Boulder, CO 80309, USA\\
\llap{$^b$}Department of Physics and Astronomy, San Francisco State University, San Francisco, CA 94132, USA\\
\llap{$^c$}RIKEN-BNL Research Center, Brookhaven National Laboratory, Upton, NY 11973, USA\\
\llap{$^d$}Raymond and Beverly Sackler School of Physics and Astronomy,
Tel Aviv University, 69978 Tel Aviv, Israel
}

\abstract{The SU(4) gauge theory with two flavors of Dirac fermions in the sextet representation shares features of a candidate for a composite Higgs model. The analogue of the Higgs multiplet of the Standard Model lives in the Goldstone manifold resulting from spontaneous breaking of the global symmetry SU(4) to SO(4). The Higgs potential arises from interaction with the particles of the Standard Model. We have computed the gauge boson contribution to the Higgs potential, using valence overlap fermions on a Wilson-clover sea. The calculation is similar to that of the electromagnetic mass splitting of the pion multiplet in QCD.}

\FullConference{34th annual International Symposium on Lattice Field Theory\\
		24--30 July 2016\\
		University of Southampton, UK}

\def\SU{{\textrm{SU}}}
\def\SO{{\textrm{SO}}}
\def\PiLR{\Pi_{LR}}
\def\CLR{ C_{LR}}
\def\eval#1{\left\langle#1\right\rangle}

\begin{document}

\section{Introduction}

UV-complete candidate theories for a composite Higgs boson \cite{Georgi:1984af} plus a partially composite top quark \cite{Kaplan:1991dc} have been catalogued by Ferretti and Karateev \cite{Ferretti:2013kya}.
Ferretti \cite{Ferretti:2014qta} has made a case for the theory built on an SU(4) gauge theory with a certain fermion content (and see \cite{Ferretti:2016upr}).
We focus here on the composite-Higgs side of the model, which is based on
a low-energy theory that is an $\SU(5)/\SO(5)$ sigma model.
This scheme of spontaneous symmetry breaking emerges from an SU(4) gauge field coupled to 5 Majorana fermions in a real representation of the gauge group.
In order to defer the difficulties attendant on odd numbers of fermion flavors (and Majorana fermions, at that), we have begun a study \cite{DeGrand:2015lna,DeGrand:2016htl} of the theory with 4 Majorana fermions, equivalent to two Dirac fermions.
As in Ferretti's model, we put the fermions into the (real) sextet representation of the SU(4) gauge group.
The global symmetry of this theory is SU(4), broken spontaneously to SO(4).

The composite-Higgs paradigm constructs the Higgs multiplet of the Standard Model from the Nambu--Goldstone bosons of the gauge theory.
The Higgs potential has its origin in loop corrections from coupling these NG bosons to the Standard Model.
I will present here a calculation \cite{DeGrand:2016htl} of the radiative part of the Higgs potential, via the vacuum polarization $\PiLR(q^2)$.
This calculation has much in common with the electromagnetic contribution to the masses of the pseudo-NG bosons of QCD, namely, the pions.
The latter is a classic calculation \cite{Das:1967it}, which has been implemented on the lattice  \cite{Gupta:1984tb,Shintani:2008qe,Boyle:2009xi}.

\section{Composite Higgs}

The aim of a composite Higgs theory is a natural construction of a Higgs boson that is protected from high energy scales.
One posits a new strong sector, called {\em hypercolor}, with a scale $f$; to hide the sector from experiments, one assumes $f\gg v$, where $v$ is the Higgs vev that defines the electroweak scale.
The hypercolor theory has spontaneous symmetry breaking, yielding a number of NG bosons.
Among these NG bosons are the Higgs multiplet $h$ of the electroweak theory.
As NG bosons, the $h$ fields are exactly massless and there is no potential $V(h)$ at all.

Once the hypercolor theory is coupled to the Standard Model, loop diagrams will generate an effective potential,
\begin{equation}
V_{\rm eff}(h)= (\alpha-4\beta)(h/f)^2 + O(h^4).
\end{equation}
The coefficient $\alpha$ has its origin in gauge boson loops,
$\alpha=(3g^2+g^{\prime2})\CLR$, where $\CLR$ is a low-energy constant that is positive definite \cite{Witten:1983ut}.
This is the subject of today's talk.
The coefficient $\beta$ comes from a top-quark loop,
$\beta=-(y_t^2/2)C_{\rm top}$, and it is probably positive; I will have no more to say about it today.
The success of the composite-Higgs model hinges on demonstrating the inequality
$4\beta>\alpha$.
This destabilizes the minimum of $V_{\rm eff}(h)$ at $h=0$ and gives the correct Higgs phenomenon of the electroweak theory.
Moreover, the induced electroweak scale had better be small, $v=\sqrt2 \eval{h} \ll f$.

\section{The model}
The low-energy theory of the hypercolor model must include the Higgs multiplet among its NG bosons.
An economical scheme is to have a global SU(5) symmetry broken spontaneously to SO(5):
Since $\SO(5) \supset [\SU(2)_L\times \SU(2)_R]$, the electroweak theory can gauge the unbroken $\SU(2)_L\times \textrm{U}(1)$ and have a custodial $\SU(2)_R$ left over.
Such a symmetry-breaking scheme demands that the hyperfermions come in a real representation of hypercolor, which restricts the hypercolor group severely if it is to be asymptotically free.
The solution%
\footnote{Ferretti \cite{Ferretti:2014qta} noted a bonus: The addition of fermions in the fundamental representation of SU(4) offers a construction of a partially composite top quark.
W.~Jay will describe this in the next talk \cite{Will}.}
 is an SU(4) gauge theory with fermions in the antisymmetric two-index representation---the sextet of SU(4).
With $N_f$ Dirac flavors we would have $\SU(2N_f)\to\SO(2N_f)$, but with 5 Majorana fermions we can have $\SU(5)\to\SO(5)$ as desired.
As I said in the introduction, for technical reasons our lattice model is not exactly the above, but close:
$N_f=2$ Dirac fermions in the sextet of hypercolor, which gives the scheme $\SU(4)\to\SO(4)$.

\section{The Higgs potential}

The gauge contribution to the Higgs potential comes from a vacuum polarization diagram (actually a difference between $VV$ and $AA$ diagrams),
\begin{equation}
 \CLR = \int_0^\infty dq^2 q^2\, \PiLR(q^2),
 \label{eq:CLR}
\end{equation}
where $\PiLR$ is a correlation function of chiral currents,
\begin{equation}
\frac12\delta_{ab}(q^2 \delta_{\mu\nu}-q_\mu q_\nu)\, \Pi_{LR}(q^2) \rule{0ex}{3ex}
  = -\int d^4x\, e^{iqx} \eval{J_{\mu a}^L(x) J_{\nu b}^R(0)}.
\end{equation}
For an analytical guess, we can try saturating the correlator with lowest resonances---the {\em minimal hadron approximation\/} (MHA),
\begin{equation}
\Pi_{LR}(q^2)\approx\frac{f_\pi^2}{q^2}-\frac{f_\rho^2}{q^2+m_\rho^2}+\frac{f_{a_1}^2}{q^2+m_{a_1}^2}.
\label{eq:MHA}
\end{equation}
This is not a bad approximation in the QCD case.

\section{Lattice calculation}

Our lattice simulations \cite{DeGrand:2016htl} used different fermion schemes for the sea fermions and the valence fermions.
We generated configurations with Wilson--clover fermions, smoothed with nHYP smearing, with an added pure gauge term to suppress nHYP dislocations.
We generated two ensembles, with different couplings but with similar lattice spacings, in order to gauge sensitivity to the lattice action without actually taking a continuum limit.
The chiral currents are sensitive to the treatment of chiral symmetry, and hence we calculated the current correlators from propagators of overlap fermions with a range of valence masses $m_v$.
In the limit $m_v\to0$, the currents satisfy exact chiral Ward identities and the correlator gives $\CLR$ via Eq.~(\ref{eq:CLR}).
The lattice volume in each case was $12^3\times24$, not large enough to eliminate finite-volume effects with any confidence.

We adopted and compared two approaches to calculating the integral $\CLR$ and taking the chiral limit.
A key difficulty is indicated by the MHA, Eq.~(\ref{eq:MHA}): $\PiLR$ has an integrable pole at $q^2=0$.

\subsection{Direct summation}

For a direct approach, we calculated $\PiLR(q_\mu)$ as a Fourier transform of the current correlator at each nonzero lattice momentum $q_\mu$.
In the summation,
\begin{equation}
C_{LR}(m_v)=\frac{16\pi^2}V \sum_{q_\mu}\Pi_{LR}(q_\mu),
\end{equation}
we include a term at $q=0$, modeling it by an integral of the continuum expression for the pole,
\begin{equation}
\Pi_{LR}(q_\mu)\simeq p+\frac{f_\pi^2}{q^2}.
\end{equation}
Here the presence of $f_\pi$ is suggested by the MHA, Eq.~(\ref{eq:MHA}); we obtain $f_\pi$ from overlap spectroscopy with the same valence mass $m_v$.
The pedestal $p$ is estimated from the values of $\PiLR$ at neighboring, nonzero momenta.
We thus obtain the result for $C_{LR}(m_v)$ plotted in Fig.~1(a).
Recall that we need the extrapolation to the chiral limit, $m_v=0$.
%%%%%%%%%%%%%%%%%%%%%%%%%%%%%%%%%%%%%%%%%%%%%%%%%%%%%%%%%%%%%%%%%%%%%%%%%%%%
\begin{figure}
\begin{center}
\includegraphics[width=.48\columnwidth,clip]{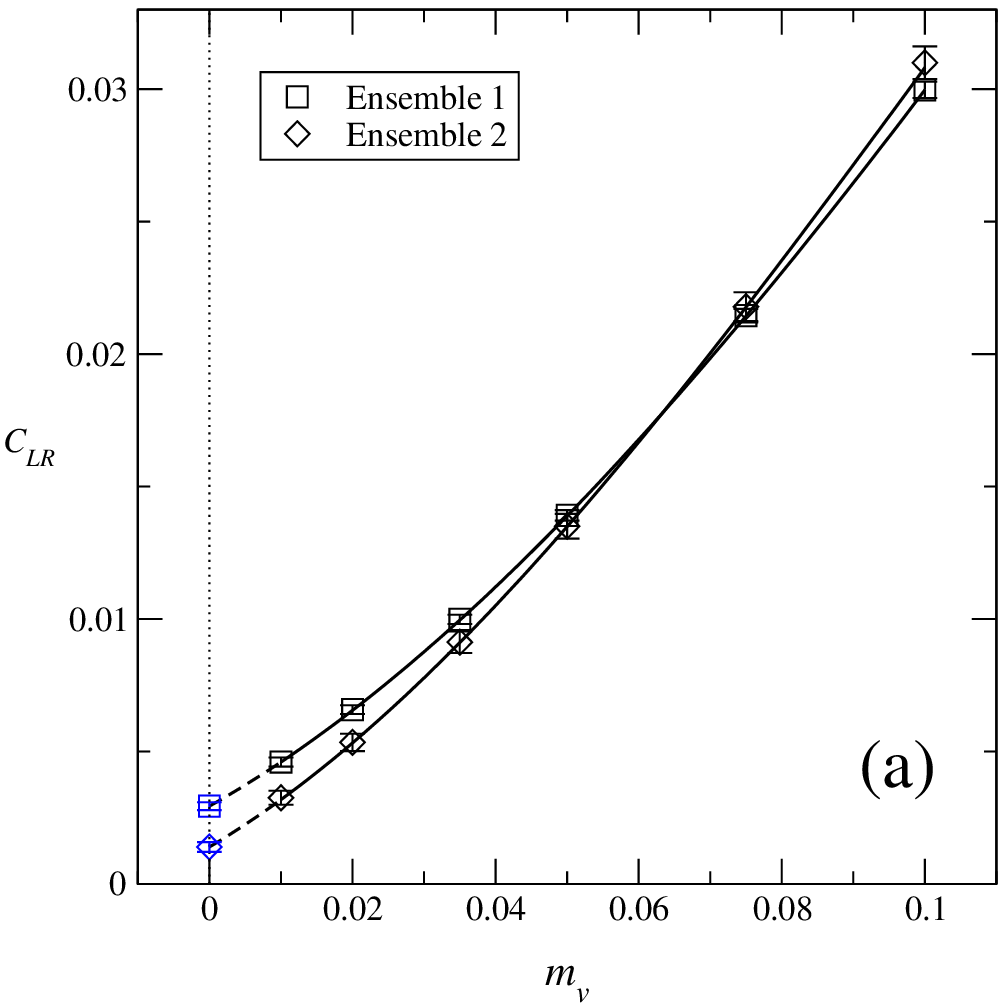}\hskip 3ex
\includegraphics[width=.48\columnwidth,clip]{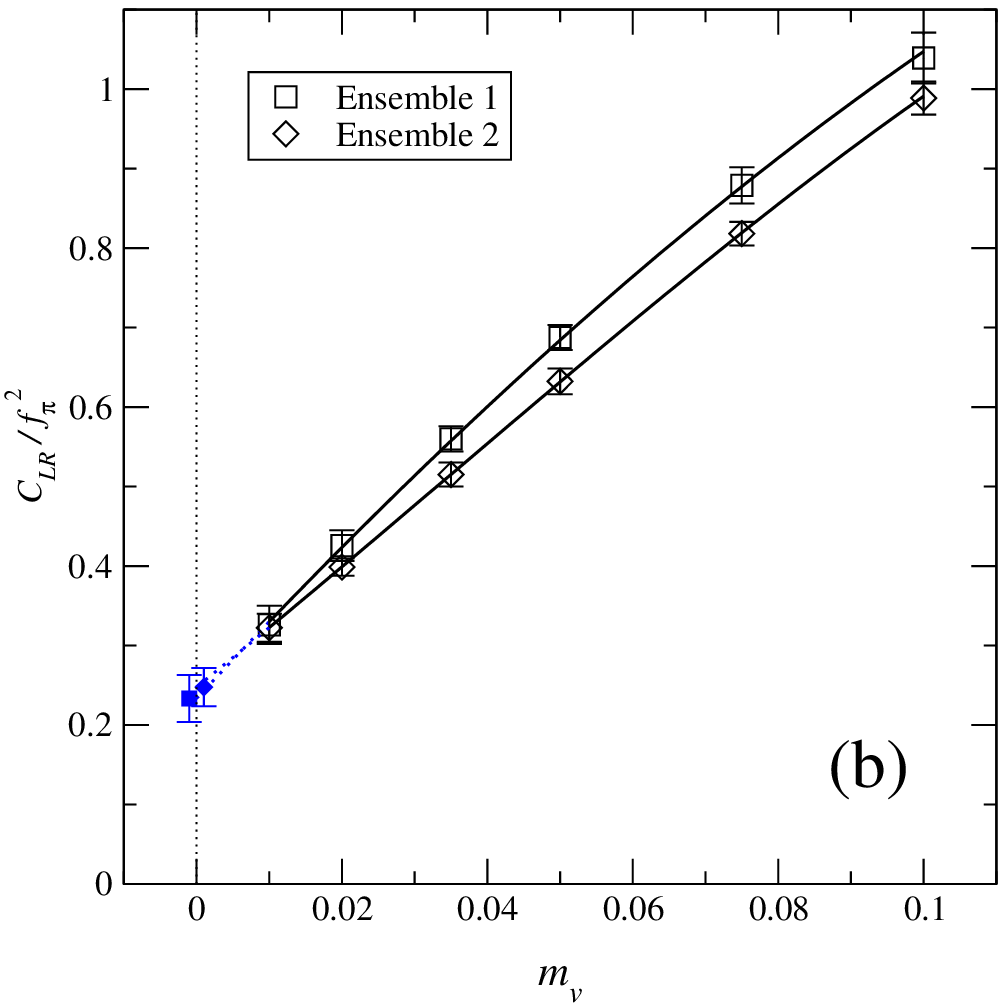}
\end{center}
\caption{(a) The low energy constant $\CLR$ as a function of valence mass $m_v$, extrapolated to zero.
The two ensembles give different results as $m_v\to0$.
(b) The ratio $\CLR/f_\pi^2$, wherein the ensembles agree.
All quantities are in lattice units.
Our two ensembles have equal lattice spacings, defined by the Sommer scale $r_1$, so the comparisons shown are valid.}
\end{figure}
%%%%%%%%%%%%%%%%%%%%%%%%%%%%%%%%%%%%%%%%%%%%%%%%%%%%%%%%%%%%%%%%%%%%%%%%%%%%

There is a clear discrepancy between the two ensembles, amounting to a factor of 2.
The two ensembles also give differing results for the valence spectra, particularly at small $m_v$.
We can show that these discrepancies have a common origin.
The integral of the MHA, Eq.~(\ref{eq:MHA}), gives
\begin{equation}
\CLR \approx f_\pi^2\, \frac{m_{a_1}^2 m_\rho^2}{m_{a_1}^2-m_\rho^2}\,
  \log\left(\frac{m_{a_1}^2}{ m_\rho^2}\right),
\end{equation}
suggesting a direct correlation between $\CLR$ and $f_\pi$.%
\footnote{Here we have used the Weinberg sum rules, valid only in the chiral limit, to eliminate $f_\rho$ and $f_{a_1}$--- even for $m_v>0$.}
Indeed, plotting the ratio $\CLR/f_\pi^2$ gives Fig.~1(b), in which we see agreement in the ratio, in the chiral limit.

\subsection{One-dimensional fit to the minimal hadron approximation}

An alternative path to $\CLR$ is to fit the data for $\PiLR(q_\mu)$ to the MHA formula,
Eq.~(\ref{eq:MHA}), treating all five constants as free parameters.
This gives a function of $q^2$ that can then be integrated numerically or analytically.
For a given value of $m_v$, the complete set of 4d data for $\PiLR$ cannot be fit to five parameters, and thus we choose a ray in momentum space, average $\PiLR$ on that ray over lattice symmetry operations, and fit to the MHA as a function of $q^2$.
Fig.~2 shows the quality of the fits, for a ray chosen along the time axis and for a space--space diagonal ray (the lattices are asymmetric in time vs.~space).
%%%%%%%%%%%%%%%%%%%%%%%%%%%%%%%%%%%%%%%%%%%%%%%%%%%%%%%%%%%%%%%%%%%%%%%%%%%%
\begin{figure}
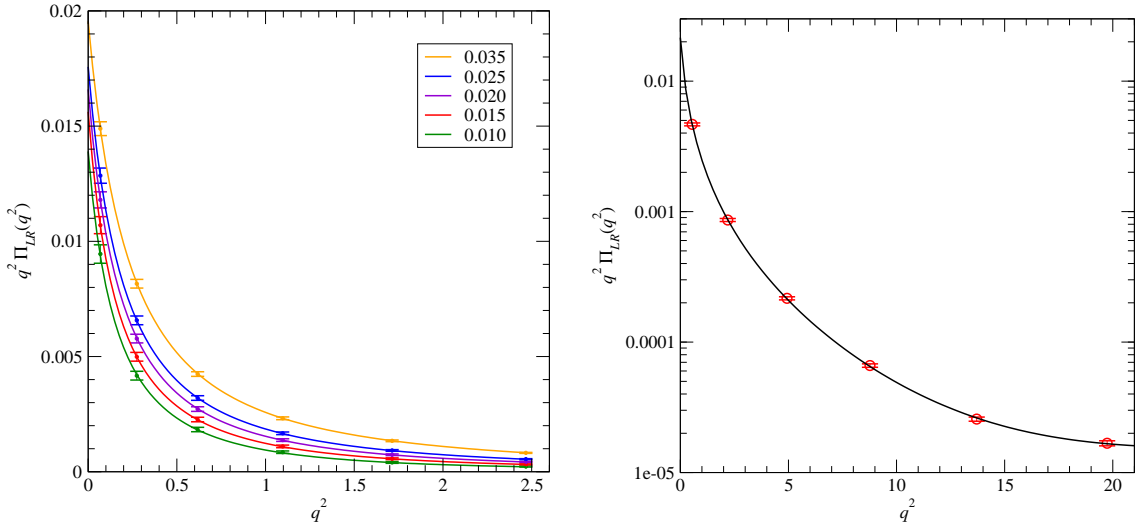

\begin{center}
\includegraphics[width=.48\columnwidth,clip]{fig4.eps}\hskip 3ex
\includegraphics[width=.48\columnwidth,clip]{fig9.eps}
\end{center}
\caption{Left: $q^2\PiLR(q^2)$ along the time axis in momentum space, fitted to the MHA formula, for 5 values of $m_v$, in ensemble 1.
Right: The same, for a single value of $m_v$, along a diagonal ray.}
\end{figure}
%%%%%%%%%%%%%%%%%%%%%%%%%%%%%%%%%%%%%%%%%%%%%%%%%%%%%%%%%%%%%%%%%%%%%%%%%%%%
In general we find that the fits have excellent $\chi^2$ as long as $m_v$ is not too large.
The parameter $f_\pi$ that emerges from the fits is consistent with the value extracted from overlap spectroscopy (and used to estimate the pole in Sec.~5.1).
The values of $m_\rho$ and $m_{a_1}$, however, do not agree well with the spectroscopic values; moreover, the corresponding decay constants come with large error bars.
This is not too surprising, since after all the MHA tries to model a multiparticle cut (in the chiral limit) with poles, and it would be surprising indeed if the parameters coincide with physical particles.

Integrating the fitted functions as in Eq.~(\ref{eq:CLR}) gives the results plotted in Fig.~3.
%%%%%%%%%%%%%%%%%%%%%%%%%%%%%%%%%%%%%%%%%%%%%%%%%%%%%%%%%%%%%%%%%%%%%%%%%%%%
\begin{figure}
\begin{center}
\includegraphics[width=.6\columnwidth,clip]{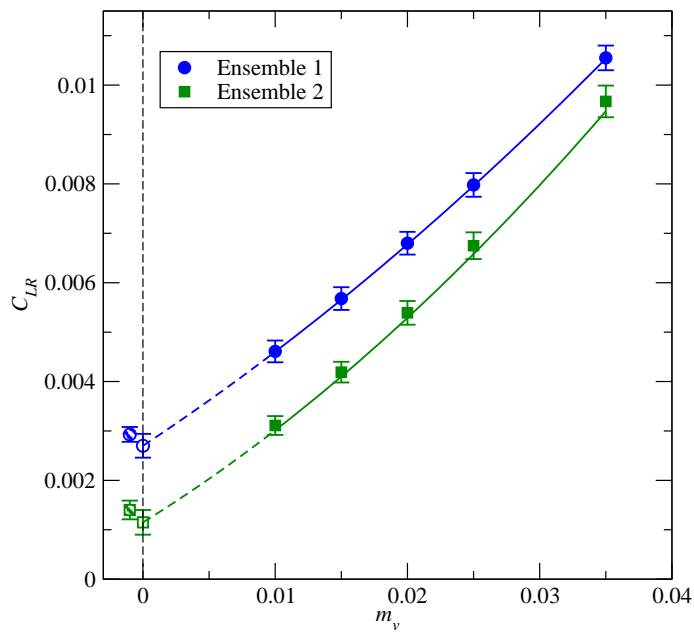}
\end{center}
\caption{The integral (\protect\ref{eq:CLR}) of $\PiLR$ obtained from fits to the MHA.
The extrapolations to $m_v=0$ agree well with the results of Sec.~5.1.
(The hashed points are the extrapolations shown in Fig.~1.)}
\end{figure}
%%%%%%%%%%%%%%%%%%%%%%%%%%%%%%%%%%%%%%%%%%%%%%%%%%%%%%%%%%%%%%%%%%%%%%%%%%%%
The extrapolations to $m_v=0$ agree well with the results of direct summation, which means, again, that the two ensembles agree when we compare values of $\CLR/f_\pi^2$.

\section{Conclusions}

We have calculated the low-energy constant $\CLR$ on two gauge ensembles, using two different procedures made necessary by the pole in $\PiLR$ at $q=0$.
We find good agreement between the methods, even though the ensembles do not agree with each other.

We have studied the systematics of the fitting and integration procedures at length, and have not found variations larger than the statistical errors plotted.
We can thus carry these procedures forward to larger volumes and towards a systematic continuum limit; we would also have to take the chiral limit in the sea quarks, according to mixed-action chiral perturbation theory \cite{Bar:2002nr,Bar:2003mh}.
While the ensembles considered here contained only sextet fermions, we will soon be applying these lessons to a model with both sextet and fundamental fermions, aiming eventually at a full implementation of Ferretti's model, with 5 of the former (Majorana flavors) and 3 of the latter (Dirac).

\smallskip

This work was supported by the U.~S. Department of Energy, Office of Science, Office of High Energy Physics, under Award Number DE-SC0010005 (T.~D. and E.~N.) and Number DE-FG03-92ER40711 (M.~G.). This work was also supported in part by the Israel Science Foundation under grant no.~449/13. Brookhaven National Laboratory is supported by the U.~S. Department of Energy under contract DE-SC0012704.

%%%%%%%%%%%%%%%%%%%%%%%%%%%%%%%%%%%%%%%%%%%%%%%%%%%%%%%%%%%%%%%%%%%%%%%%%%%%
%\newpage

\end{document}